\documentclass[review]{elsarticle}

\usepackage{lineno,hyperref}
\modulolinenumbers[5]

\journal{Journal of Reliability Engineering and System Safety}

\usepackage{comment}
\usepackage{lineno}

\usepackage{algorithm}
\usepackage{algorithmic}

\usepackage{amsfonts}

\usepackage{makecell}

\usepackage{appendix}

\usepackage{amsmath} 
\DeclareMathOperator*{\argmax}{arg\,max}

\begin{document}

\begin{frontmatter}

\title{System Effects in Identifying Risk-Optimal Data Requirements for Digital Twins of Structures}



\author[address_1,address_2]{Domenic Di Francesco}

\author[address_2]{Max Langtry}
\author[address_1,address_3]{Andrew Duncan}
\author[address_1,address_4]{Chris Dent}


\address[address_1]{The Alan Turing Institute for Artificial Intelligence and Data Science, The British Library, 2QR, John Dodson House, 96 Euston Rd, London NW1 2DB}

\address[address_2]{Department of Civil Engineering, Cambridge University, Trumpington Street, CB2 1PZ}

\address[address_3]{Department of Mathematics, Imperial College London, Huxley Building, South Kensington Campus, London, SW7 2AZ}

\address[address_4]{Department of Mathematics, University of Edinburgh, James Clerk Maxwell Building, Peter Guthrie Tait Road, Edinburgh, EH9 3FD}

\begin{abstract}




Structural Health Monitoring (SHM) technologies offer much promise to the risk management of the built environment, and they are therefore an active area of research. However, information regarding material properties, such as toughness and strength is instead measured in destructive lab tests. Similarly, the presence of geometrical anomalies is more commonly detected and sized by inspection. Therefore, a risk-optimal combination should be sought, acknowledging that different scenarios will be associated with different data requirements.

Value of Information (VoI) analysis is an established statistical framework for quantifying the expected benefit of a prospective data collection activity. In this paper the expected value of various combinations of inspection, SHM and testing are quantified, in the context of supporting risk management of a location of stress concentration in a railway bridge. The Julia code for this analysis (probabilistic models and influence diagrams) is made available. The system-level results differ from a simple linear sum of marginal VoI estimates, i.e. the expected value of collecting data from SHM and inspection together is not equal to the expected value of SHM data plus the expected value of inspection data. In summary, \textbf{system-level decision making, requires system-level models}.


\end{abstract}


\end{frontmatter}

\linenumbers

\begin{table}
    \centering
    \begin{tabular}{c|c|c}
        Symbol & Meaning & Units\\
        \hline
        $a$ & risk mitigation action & $-$ \\
        $a^{*}$ & risk mitigation action associated with expected optimal utility & $-$ \\
        $A$ & set of available risk mitigation action(s) & $-$ \\
        $C_{fail}$ & cost of failure & $-$ \\
        $C_{rm}$ & cost of risk mitigation & $-$ \\
        $e$ & data collection activity & $-$ \\
        $e^{*}$ & data collection activity associated with expected optimal utility & $-$ \\
        $E$ & set of available data collection activities & $-$ \\
        $m$ & mean & $-$ \\
        $\Pr(fail)$ & probability of failure & $-$ \\
        $sd$ & standard deviation & $-$ \\
        $u$ & utility & $-$ \\
        $Var$ & variance & $-$ \\
        $z$ & measurement data & $-$ \\
        $f_{X}(x)$ & probability mass or density function for uncertain parameter, $x$ & $-$ \\
        $\alpha$ & shape parameter of Gamma distribution & $-$ \\
        $\beta_{R}$ & reliability index & $-$ \\
        $\gamma$ & scale parameter of Gamma distribution & $-$ \\
        $\delta$ & misalignment & $mm$ \\
        $\epsilon$ & measurement uncertainty parameter & $-$ \\
        $\theta$ & uncertain and unobserved model parameter(s) & $-$ \\
        $\rho$ & correlation coefficient & $-$ \\
        $\sigma_{L}$ & applied stress & $MPa$ \\
        $\sigma_{L-meas}$ & stress inferred from sensor data & $MPa$ \\
        $\sigma_{Y}$ & yield strength & $MPa$ \\
        $\Sigma$ & covariance matrix & $-$ \\
        $\Phi_{G}$ & \makecell{multivariate distribution defined by a Gaussian copula, \\associated covariance matrix and marginal distributions} & $-$ \\
    \end{tabular}
    \caption{Nomenclature}
    \label{tab:nomenclature}
\end{table}

\newpage
\nolinenumbers
\section{Introduction} \label{sect:into}

\subsection{Risk Based Structural Integrity Management} \label{sect:intro_risk}

Risk is defined as the expected consequences of uncertain outcomes \cite{Faber2012}, and can therefore be used to rank decision alternatives consistently with statistical decision theory \cite{VonNeumann1953}. For instance, it is expected to be worthwhile investing in a maintenance activity if it is believed that the corresponding net reduction in failure costs is greater than the costs required to complete the work. By the same approach, the optimal activity (where multiple may be available, including the option to take no action) is that for which this net benefit is the greatest.

In industrial Risk Based Inspection (RBI) schemes, calculation complexity is often cited as the reason that simplified heuristics are used, rather than the above-mentioned principled statistical decision analysis \cite{AmericanPetroleumInstitute2016, DNVGL2017d}. Due to the emergence of novel data collection technologies and methods for scalable probabilistic inference, it can be argued that these justifications for avoiding quantitative statistical methods are now less valid \cite{DiFrancesco}. The principle of RBI is that resource allocation should be directed by risk, i.e. as components in a system become increasingly likely to fail, or the associated failures become increasingly costly (or both), their priority in maintenance budgeting should also increase. Without an absolute scale of risk (which can only be obtained from a fully quantitative risk analysis), it cannot be determined for which components investment is expected to be worthwhile, only their relative priorities can be identified, and not always consistently. Simple rules will always have pragmatic and implementation benefits, but can lead to sub-optimal resource allocation and excessive (unquantified) risk. The barriers associated with the introduction of methods from computational statistics will decrease as the cost of formulating and performing calculations decreases and training in data-centric engineering becomes widespread. The associated benefits of transparent, quantitative and auditable decision-making are then expected to emerge.

\subsection{System Effects} \label{sect:intro_sys}

Recent technological developments in data collection for engineering structures have largely focused on Structural Health Monitoring (SHM). This can broadly be described as the process of fitting sensors to structural components (either during construction of new builds, or as retrofits to existing structures) so that local conditions can be measured in operation \cite{Neves2020}. A typical application is that of installing strain gauges to better understand the applied loads at particular locations. Alongside developments in data analysis, across many industries, this may have a transformative impact on the risk management of much of the built environment.

The expected value of this data, which can be quantified in the context of a decision analysis (see Section \ref{sect:voi}), may be especially high when there is little other information available regarding the local environmental conditions. However, in instances where these are considered to be adequately characterised in prior models, it may not be worthwhile installing SHM systems. Rather, one of the other features of a structural integrity assessment may benefit from measurement.

\begin{figure} \centering
    \includegraphics{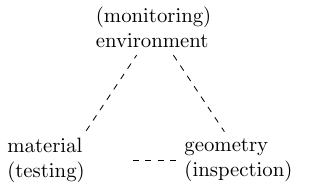}
    \caption{Features of an assessment of structural condition}
    \label{fig:ECA_triangle}
\end{figure}

\begin{figure} \centering
    \includegraphics{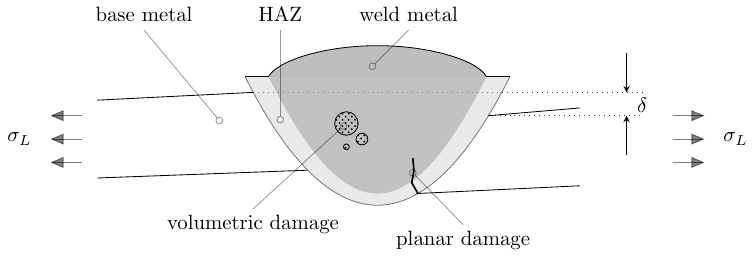}
    \caption{Geometric features of welded steel connection, including plate misalignment, $\delta$}
    \label{fig:geometry}
\end{figure}

It is therefore important to recognise that the value of SHM data may be influenced by the present understanding of the wider structural system, which can be modelled by integrating multiple sources of information. A simple example is shown in Figure \ref{fig:ECA_triangle}, commonly referred to as the Engineering Critical Assessment (ECA) of fracture mechanics triangle \cite{Anderson2005}, which describes the key sources of information in an assessment of structural condition. As an example, consider the diagram in Figure \ref{fig:geometry} of two plates (base metal) joined by a weld consisting of weld metal and a Heat Affected Zone (HAZ), which contains anomalies. Here, data from SHM may provide an indirect and imperfect measurement of the applied stress, $\sigma_{L}$. Inspection, often referred to as Non-Destructive Testing/Evaluation (NDT/NDE), can provide similarly imperfect detection and sizing of the presence/extent of damage\footnote{Note that differing technologies are better suited to different types of damage (see Annex T in \cite{BSI2015})}. Finally, material testing can provide a better understanding of the properties of the weld, HAZ, and base metal. All of these remain important inputs in empirical fracture mechanics models \cite{DiFrancesco2022}. As SHM technologies develop, it may be possible to reliably infer material and geometrical properties from the data they provide. The methods and calculations discussed in this paper are, in principle, agnostic to the source of the data. Rather, they are concerned with a mathematical characterisation of the quality of data required by a model, in the context of the underlying decision problem for which the model is required, see Section \ref{sect:voi}.

By considering the inter-dependencies between multiple sources of information, engineers can better assess the overall health of a structure and make more principled risk management decisions.

\subsection{Probabilistic Digital Twins} \label{sect:intro_pdts}

There are various, and often contradictory definitions of the term \emph{digital twin}. A UK government report \cite{Bolton2018} proposes the following definition, which is considered suitably broad: 
\begin{quote}
    Digital twins are realistic digital representations of physical things. They unlock value by enabling improved insights that support better decisions, leading to better outcomes in the physical world.
\end{quote}

The calculations and discussions in this paper are regarding how mathematical representations of decision problems can be used to obtain transparent and replicable (auditable) risk management strategies for structures. Such analysis is conditional on an underlying model, which may be considered to be a digital twin of the structure. Note that differing requirements have been specified for what constitutes a \emph{digital twin}, such as the requirement for a 3-dimensional model, or for at least one source of data to be in the form of a (near) real-time stream. 

In practice, the form and constituent components of a digital twin (or any engineering model) should be based on the requirements of the underlying decision problem that the model is intended to support. The decision to invest in a sensing system should be based on the extent to which the data is expected to facilitate improved risk management (see Section \ref{sect:voi}), and not so that the model can be categorised as a digital twin\footnote{Unless there are additional benefits to this, in which case they should be made explicit in the model and accounted for.}. Engineers have a legal \cite{Harris2003} and ethical \cite{TheRoyalAcademyofEngineering2011} responsibility to develop and maintain safe systems, and this link between underlying risk management decisions and structural calculations/analysis should always be clearly defined, as it will be scrutinised following a failure \cite{Hopkins2002}. Digital twins should be developed to effectively support decisions subject to uncertainty, such as what quality, and quantity of data is required for a structure.

\section{Decision Analysis} \label{sect:voi}

\subsection{Conceptual Introduction} \label{sect:voi_formal}

The need to support high consequence decisions is what dictates requirements for various engineering analysis. For instance, material selection during a design may be informed by stress analysis, or maintenance investments can be informed by fracture mechanics assessments. However, unless the purpose is made explicit, the results of such calculations may require some intermediary interpretation before they can be used to support a decision. This means that there is an opportunity to introduce unquantified and undocumented bias into the analysis.

Influence diagrams \cite{Howard2005} offer a concise graphical representation of the components of a decision problem, namely uncertainties, costs and possible actions. Domain knowledge is needed to design influence diagrams, as the causal model used effects the results of any calculations performed, with the addition or removal of arrows potentially leading to entirely different outcomes \cite{Pearl2016}.

\begin{figure} \centering
  \includegraphics{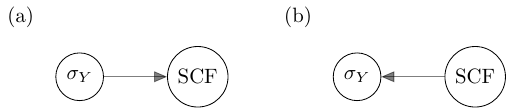}
  \caption{DAGS (a) and (b) conveying alternative causal explanations for an association between $\sigma_{Y}$ and SCF}
  \label{fig:alternative_casual_models}
\end{figure}

\begin{figure} \centering
    \includegraphics{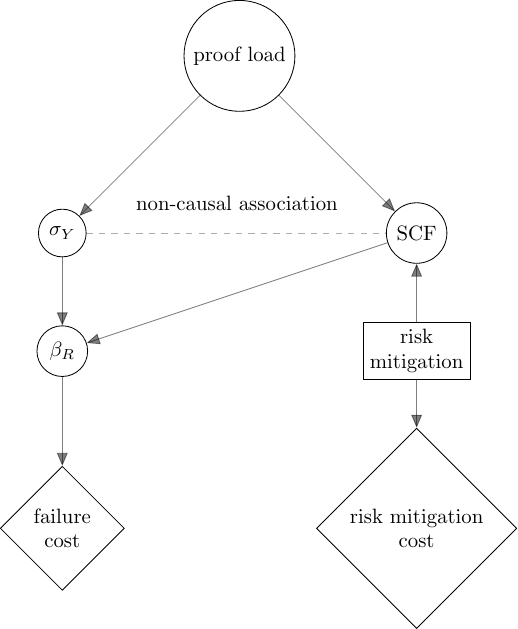}
    \caption{Influence diagram proposing non-causal association between $\sigma_{Y}$ and SCF}
    \label{fig:non_causal_association}
\end{figure}

As an example, consider the case of an inspection for geometric anomalies on a structure. If predicted to be associated with sufficiently large stress concentration effects (characterised by a Stress Concentration Factor, SCF), the relevant components were replaced and tensile testing specimens machined from them. When compared with pre-commissioning test data, the yield strength, $\sigma_{Y}$ was found to be relatively high at these locations of high SCF. This association can be described in various ways. In Figure \ref{fig:alternative_casual_models}, diagram (a) proposes that a high strength causes a high SCF, and diagram (b) proposes that a high SCF causes a high strength. A more suitable causal model identifies that the proof load that the structure was exposed to is a confounding variable\footnote{A simple explanation for this would be that, knowing a component has been exposed to a certain load provides evidence for: the stresses and the extent of damage present at the time. A large load at the presence of a high SCF would require a relatively high strength to not fail.}, as shown in Figure \ref{fig:non_causal_association}. The reason that this representation matters, is because it will impact risk management decisions. If a modelling team erroneously considered diagram (b) in Figure \ref{fig:alternative_casual_models}, then any interventions that impact the SCF would also be assumed to change the material strength. Domain expertise is required to design the structure of influence diagrams, and prior models. However, they can then be used to find associated expected optimal decision strategies.

Implications of such features in DAGs are discussed in more detail in \cite{McElreath2019} from a causal inference perspective, and in \cite{Glavind2021} with reference to engineering applications.

As suggested in Section \ref{sect:intro_pdts}, deciding whether or not to invest resources in collecting and analysing data is an on-going challenge in engineering. Fundamentally, this requires an evaluation of the question:

\emph{how} and \emph{to what extent} is the proposed data expected to facilitate improved risk management?

Value of Information (VoI) analysis provides a replicable and quantitative framework for addressing this question, as demonstrated by the example calculations in this paper, and in recent scientific literature \cite{Malings2016, Straub2017, Thons2018, DiFrancesco}. It achieves this by considering the engineering models and the underlying decision problem jointly. This helps target analysis in various ways, including identifying when and where purchasing data of a specified quality is expected to be worthwhile.

Applying VoI to structural integrity management raises the question, "what is it about data that provides value?" The mechanism is that data reduces statistical (epistemic) uncertainty, for example, as more tensile tests are completed, engineers can better understand the strength of a material. 

As the uncertainty reduces, decision problems become simpler\footnote{Here, \emph{become simpler} means that there is less uncertainty associated with an estimate of an expected optimal decision. In the extreme (where there is no uncertainty remaining) it is possible to simply compare known outcomes.} and decision makers may identify new expected optimal strategies, or simply expose themselves to less risk as a consequence of more precisely quantifying characteristics of their system. 

A key challenge of VoI analysis is that the value that data will eventually provide to a decision maker, will depend on the result of the measurement(s) obtained. However, this needs to be evaluated before the measurement is taken. This challenge is common to Bayesian experimental design \cite{Chaloner1995}, which is also concerned with optimising data collection. The solution is to use a prior model that describes the uncertainty in the quantity of interest. The calculation compares the risk if a decision maker were to act based on this prior model, with the expected risk when some plausible data is incorporated. This comparison answers the initial question of how and to what extent data is expected to improve risk management decision making. When this difference is quantified on a monetary scale (as is typically the case in VoI analysis), it can be interpreted as how much an engineering team should be willing to pay for the data.

Some important features of VoI analysis are summarised below:
\begin{itemize}
    \item \textbf{Challenge in generalising results}. The expected value of data in the context of solving a decision problem is case-specific. Changing a single element of the problem may or may not change the results, and this outcome may not be intuitive.
    \item \textbf{Quantity vs. Quality trade-off}. For instance, for a given bridge, it may be more worthwhile to collect small amounts of load data than large amounts of high-resolution displacement data. Similarly the value of load measurements may vary with location over the span.
    \item \textbf{Data is imperfect}. Considering perfect information provides a useful upper bound and simplifies calculations. Accounting for the imperfect features (precision, bias, reliability, missingness) of data in the likelihood function produces more realistic results at greater computational expense.
\end{itemize}

\subsection{Formal Definitions} \label{sect:voi_math}

\begin{figure} \centering
    \includegraphics{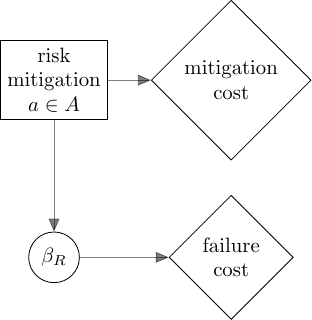}
    \caption{Generic influence diagram for mitigating risk of structural failure}
    \label{fig:id_generic}
\end{figure}

Consider the influence diagram in Figure \ref{fig:id_generic}. Here, the structural reliability of a system $\beta_{R}$ is defined based on some set of uncertain parameters, $\theta$. It determines the expected consequence of failure (along with a cost of failure). A set of risk mitigation options, $A$, which include the option to take no action, are available to the operator. The selection of this action is required to find the cost of risk mitigation. The optimal action, $a^{*}$, is defined as that which maximises the expected utility (or, equivalently, minimises the expected cost), as defined in Equation \ref{eq:prior_decision} \cite{Pratt1995, Faber2012}.

\begin{equation}
    a^{*} = \argmax_{a \in \mathcal{A}} \mathop{\mathbb{E}}_{\theta \sim f_{\Theta}(\theta)} \big[ u(a,\theta) \big]
    \label{eq:prior_decision}
\end{equation}

\begin{figure} \centering
    \includegraphics{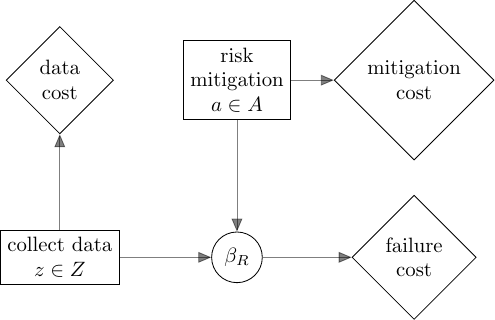}
    \caption{Generic influence diagram for quantifying the expected value of structural data}
    \label{fig:id_voi_generic}
\end{figure}

VoI analysis, originally proposed in \cite{Raiffa1961}, considers the opportunity to collect data and investigates how this would impact the decision analysis. The basic problem in Figure \ref{fig:id_generic} is extended in Figure \ref{fig:id_voi_generic} to introduce data collection opportunities, $\mathcal{E}$, including the option to proceed with no additional data. Since this data, $z$, will generally be some indirect measurement of $\theta$, it will influence the estimate of $\beta_{R}$. The two decision variables (risk mitigation and data collection) can be jointly optimised, as shown in Equation \ref{eq:prepost_decision}. The expected value of a specific data collection activity, $e_{i}$, is defined as the difference between the expected utility with and without the data. This result can be interpreted as how much the operator should be willing to pay for the data, meaning that it can directly inform budgeting decisions regarding data acquisition.

\begin{equation}
    e^{*}, a^{*} = \argmax_{\substack{a \in \mathcal{A} \\e \in \mathcal{E}}} \mathop{\mathbb{E}}_{\substack{\theta \sim f_{\Theta}(\theta) \\ z \sim f_{Z}(z \mid \theta)}} \big[ u(a, e, f_{\Theta}(\theta \mid z) \big]
    \label{eq:prepost_decision}
\end{equation}

\begin{equation}
    VoI(e_{i}) = \mathop{E}_{\substack{\theta \sim f_{\Theta}(\theta), \\ z \sim f_{Z}(z \mid \theta)}} \big[ u(e_{i}, a^{*}, f_{\Theta}(\theta \mid z) \big] - \mathop{E}_{\theta \sim f_{\Theta}(\theta)} \big[ u( a^{*}, f_{\Theta}(\theta)) \big]
    \label{eq:voi}
\end{equation}

The basic calculation procedure is outlined in Algorithm \ref{alg:VoI}. The term \emph{preposterior} in this algorithm is used in engineering to note that the prior model is being updated using hypothesised data, rather than actual measurements (which are not available at the time of evaluating whether obtaining these measurements is risk-optimal) \cite{Jordaan2005}. The steps involving identifying the optimal action, and associated expected optimal utility require solving an influence diagram, i.e. finding the strategy (combination of available actions) that maximises the expected utility. In this paper, this optimisation has been solved as a mixed-integer linear program using an extension \cite{Salo2021} to the Julia mathematical optimisation library \cite{Lubin2022, Optimization2023}. This approach includes the constraint that all (decision) variables are discrete, $a \in \mathbb{Z}^{n}$, representing the binary option to implement an action or not. Each combination of actions $a \in A$ can therefore be assigned a discrete index, and the maximum expected utility over this set will be associated with $a^{*}$ \cite{Kochenderfer2019}.
    
\begin{algorithm}
\caption{Quantitative estimate of expected value of data collection}
\label{alg:VoI}
\begin{algorithmic}

    \STATE 1. Identify optimal action $a^{*}$, with respect to the expected utility, $\mathop{E}[u]$, given prior distribution(s) $f_{\Theta}(\theta)$
    
    \STATE 2. Find associated expected prior utility $E[u(a^{*})]$

    \FOR{hypothesised measurement, $z_{e_{i}}$, in $\theta \sim f_{\Theta}(\theta)$}
    
        \STATE 3a. Define likelihood function $f_{Z}(z_{e_{i}} \mid \theta)$ describing information content of data, from measurement activity $e_{i}$
        \STATE 3b. Update prior model to obtain \emph{preposterior} distribution, $f_{\Theta}(\theta \mid z_{e_{i}})$
        \STATE 3c. Identify $a^{*}$ given preposterior distribution
        \STATE 3d. Find associated expected prior utility $E[u(a^{*}, z_{e_{i}})]$
    
    \ENDFOR

    \STATE 4. Calculate expected value of proposed measurement:
    \STATE $VoI = \dfrac{1}{N samples} \times \sum_{n = 1}^{N samples} E[u(a^{*}, z_{e_{i}}(n))] - E[u(a^{*})]$
    
\end{algorithmic}
\end{algorithm}

\subsection{Practical Considerations: Qualitative Example} \label{sect:voi_imp}

A desirable feature of VoI analysis is that it is a quantitative and replicable method for identifying risk-optimal data collection activities, to the extent that the associated models provide a valid representation of the true system. For instance, some pipelines in the USA are believed to be over $50$ years old, and still in operation despite missing design documentation \cite{DepartmentofTransportationDOT2023}. There are instances where the material (grade of linepipe steel) is unknown. Although it may be of interest to retrofit SHM sensors to such a structure, a VoI analysis is likely to identify that it is expected to be more worthwhile to understand the material properties. Depending on the intended future operation, it may also be worthwhile to collect other types of data too, but as argued in this paper, there should be a transparent, risk-based justification for any measurements or interventions.

As a more detailed example, consider the Morandi cable stayed bridge, which collapsed in Genoa (Italy) in 2018, resulting in death of 43 people. As outlined in \cite{Rania2019, Clemente2020}, previous attempts to investigate the condition of the concrete had failed to obtain the required information.

\begin{figure}\centering
    \includegraphics{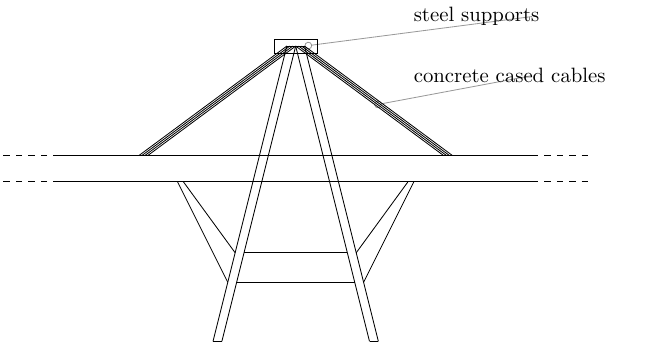}
    \caption{Schematic diagram of failed section of Morandi bridge}
    \label{fig:Morandi_diagram}
\end{figure}

The means by which data collection from the bridge could have prevented the catastrophic failure are considered below.

\begin{itemize}
    \item \textbf{Strain gauges}: measured strains may have triggered some intervention if cables failed sequentially, with sufficient time between failures to detect the redistribution of load.
    \item \textbf{Inspection of cables}: provided that the inspection was able to obtain measurements of the damage present in the cables (an earlier attempt had not, \cite{Clemente2020}), this would have identified an important factor in the reduced reliability of the bridge.
    \item \textbf{Material testing of cables and concrete}: in the failure investigation, concrete was measured to be approximately three times weaker than the expected strength. This was attributed to illegal activities during the construction of the bridge in the 1960's. Testing of samples prior to the collapse could have identified this.
\end{itemize}

Chloride-induced corrosion was known to be a threat, given the marine environment that the bridge was exposed to. Propagating uncertainty in the condition of the concrete (and perhaps also the strength of the concrete) through a decision analysis could have shown that investments in data collection and subsequent risk mitigation may have been worthwhile. Without formalising this procedure, maintenance decisions lose transparency, as different interpretations of reports, calculations and confounding variables can be used to justify any outcome.

In 2018 steel supports were added to the span of the bridge that failed, see Figure \ref{fig:Morandi_diagram}. Strengthening repairs are a valid form of risk mitigation, but will not necessarily be the best option. Such interventions should be justified using a decision analysis, with a transparent path from the intervention to the specific source of risk that is being targeted. In this case, strengthening the concrete towers did not improve the reliability of the corroding cables. Alternative options may include limiting the traffic (operational cyclic loads) on the bridge, or replacing parts with new components with more precisely known properties.

There are organisational challenges in integrating some novel methods of data collection and analysis into the business processes of the asset owner, in this case Autostrade per l'Italia. The introduction of data-driven management may improve safety by benefitting from recent research, but any existing workflows that are superceded should be carefully considered, as subject matter expertise (for instance, in defining influence diagrams and prior probabilistic models), is not presently straightforward to automate, particularly for specific assets. In this case, a generic review may have identified the threat of Chloride-induced corrosion, but a knowledge of the design and operational history would be able to confirm the extent to which this had been mitigated, or measured. The absence of useful data should have been evident in an increased uncertainty in the cable condition, which may then have identified some available risk mitigation as a worthwhile investment. These concepts are demonstrated quantitatively, in Section \ref{sect:application}.

\section{Quantitative Example: Risk Management for a Railway Bridge} \label{sect:application}

\subsection{Introduction} \label{sect:staff_bridge}

\begin{figure}
    \includegraphics[width = \textwidth]{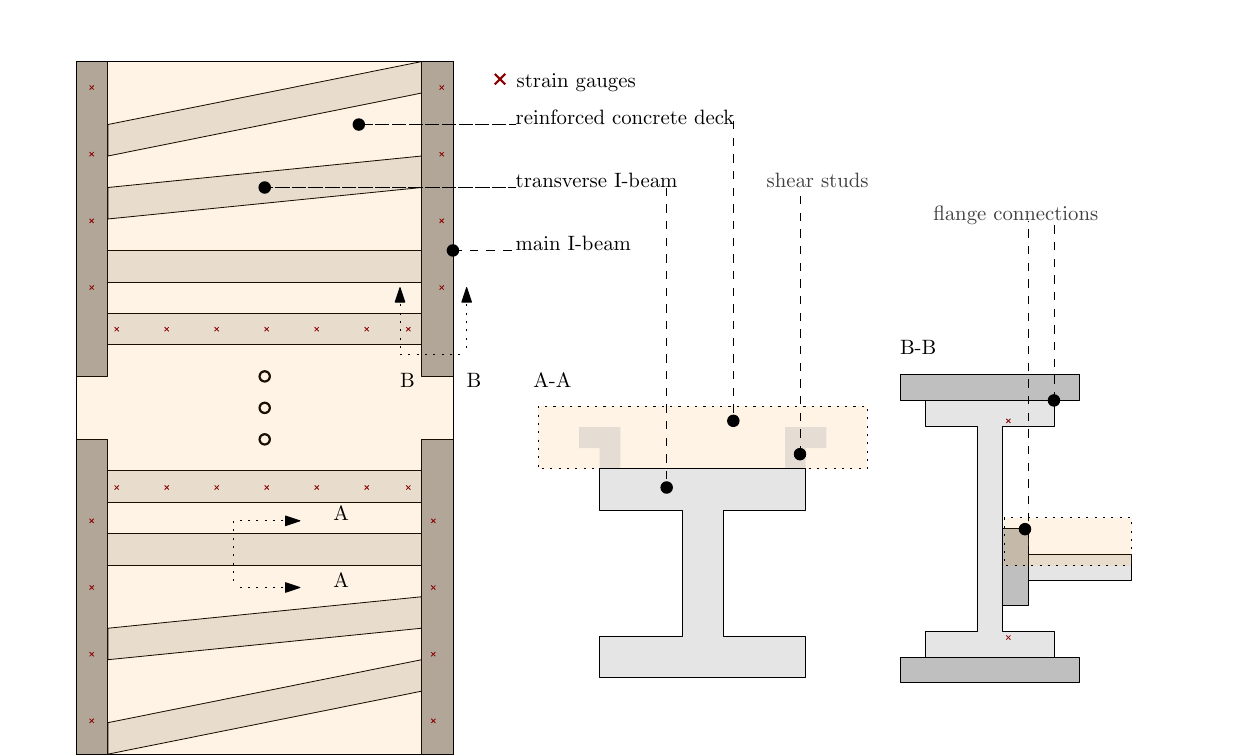}
    \caption{Schematic diagram of strain gauge placement on elements in Staffordshire railway bridge}
    \label{fig:Stafforshire_bridge}
\end{figure}

A bridge in Stafforshire, UK, was designed to carry passenger trains and was constructed with an SHM system \cite{Butler2018}. Specifically, Fibre Bragg Grating (FBG) strain gauges were installed at multiple locations, a diagram of their arrangement is shown in Figure \ref{fig:Stafforshire_bridge}. Here, these sensors were placed on the main I-beams ($20$ each), across the span of the bridge, and on two selected transverse beams ($7$ each). Tensile strains of approximately $25 \mu$ strain were recorded during a train passing \cite{Febrianto2022}. 

This bridge is used as an example to demonstrate how maintenance decisions subject to uncertainty can benefit from additional data collection. The values (e.g. material properties, loads, and costs) used in the calculation do not precisely match those for the Staffordshire rail bridge, but in Section \ref{sect:staff_bridge_calc} are argued to be representative of such a problem. 

\begin{figure}\centering
    \includegraphics{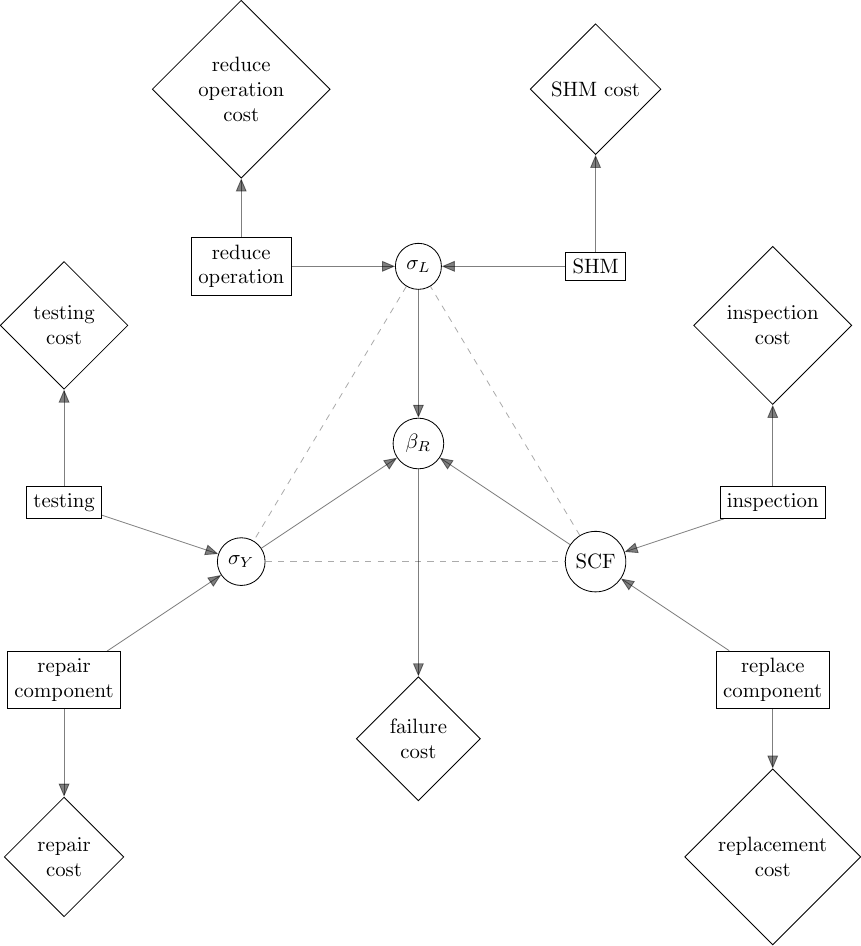}
    \caption{Extension of "ECA triangle" (see dashed lines) to influence diagram representation of structural integrity management decision problem}
    \label{fig:rail_sim_id}
\end{figure}

The maintenance decision problem is represented by the influence diagram in Figure \ref{fig:rail_sim_id}, which has been solved as part of this example. Here, each of the elements of the ECA triangle in Figure \ref{fig:ECA_triangle} is considered (as shown by the dashed lines), with options to collect data, or reduce the risk at each of the three elements. These elements then jointly inform (an uncertain estimate of) the reliability index, $\beta_{R}$. The load, and geometry in particular, may vary along the span of the bridge. The influence diagram in Figure \ref{fig:rail_sim_id}, and associated probabilistic models in Section \ref{sect:staff_bridge_calc} define the conditions at a single joint. Misaligment at flange connections, such as those in the Staffordshire railway bridge, or at welded connections, such as in Figure \ref{fig:geometry}, both concentrate stresses. 

Specifically, an upcoming annual maintenance window for the bridge is considered. The failure mechanisms considered are listed below:
\begin{itemize}
    \item \textbf{Over-stress}: This limit state is defined as the combined effect of the applied load and the stress concentration exceeding the yield strength.
    \item \textbf{Fatigue}: The repeated application of loads, which may individually be insufficient to cause failure, exceeding a permissible limit, as defined by an SN curve\footnote{This model proposes a linear relationship (on a logarithmic scale) between the amplitude of repeated stress cycles, $S$, and the number of cycles at $S$ before fatigue failure, $N$.}. Test data has been simulated from a curve for a class D joint \footnote{This category is considered representative for the hot-spot stress (at locations where cracks are more likely to initiate) of a range of welded joints and flange connections} \cite{BSI2014}, and a probabilistic model has been fit to account for the variability. See the calculation for complete details \cite{DiFrancesco2023}.
\end{itemize}

The probability of failure, $Pr(fail)$, is defined as the probability of exceeding of at least one of these limit states, see Equation \ref{eq:pr_fail_1}. This is calculated using probabilistic models of yield strength, applied stress, and stress concentration. The available actions considered in the decision problem are presented in Table \ref{tab:bridge_actions}. Note that this analysis considers that \emph{any combination} of these actions can be selected, including the option to take no action, and to collect all sources of data and implement all available risk mitigation.

\begin{equation}
    \Pr(fail) = \Pr(\textrm{over-stress} \cup \textrm{fatigue})
    \label{eq:pr_fail_1}
\end{equation}

\begin{table}[]
    \centering
    \begin{tabular}{|c|c|c|c|}
         \Xhline{2\arrayrulewidth}
         Action & Action type & Child node & Description \\
         \Xhline{2\arrayrulewidth}
         testing & data collection & $\sigma_{Y}$ & reduces uncertainty in material properties \\
         \hline
         repair & risk mitigation & $\sigma_{Y}$ & \makecell{strengthening repair increases \\capacity to withstand load. \\Stress is multiplied by $0.75$}\\ 
         \hline
         inspection & data collection & SCF & reduces uncertainty in local geometry \\
         \hline
         replace & risk mitigation & SCF & \makecell{replacing misaligned component \\reduces stress concentration effects \\(if correctly installed). \\SCF is multiplied by $0.0$}\\
         \hline
         SHM & data collection & $\sigma_{L}$ & reduces uncertainty in applied load \\
         \hline
         reduce operation & risk mitigation & $\sigma_{L}$ & \makecell{limiting the frequency of train passage \\reduces the number of stress cycles \\ experienced, increasing fatigue life.\\ Cyclic loading frequency is mulitplied by $0.5$}\\
         \hline
    \end{tabular}
    \caption{Summary of intervention options available to operator and qualitative explanation of their effect on downstream nodes}
    \label{tab:bridge_actions}
\end{table}

\subsection{Decision Problem and Utility Function} \label{sect:staff_bridge_calc}

The decision problem described by the influence diagram in Figure \ref{fig:rail_sim_id} is defined by probabilistic models of the uncertain parameters (including the effect of the various interventions, which are summarised in Table \ref{tab:bridge_actions}), and the costs of each outcome. Note that these models describe a specific element, at which some misalignment, and therefore stress concentration, is believed to be present.

The prior model for stress, $\sigma_{L}$, is shown in Equation \ref{eq:prior_stress}. The mean value of $50 MPa$ is approximately equivalent to the (elastic) stress associated with an SHM measurement of $25 \mu$ strain \cite{Febrianto2022} and an elastic modulus of $210 GPa$, as assumed in \cite{Butler2018}. A joint prior model of stress concentration factor, $SCF$, and yield strength, $\sigma_{Y}$, is presented in Equation \ref{eq:prior_SCF_Strength}. Here some dependency is proposed to account for the effect of previous proof load testing, which makes the combination of relatively low material strength and relatively high stress concentration to be less likely (see Figure \ref{fig:alternative_casual_models}).

\begin{equation}
    \sigma_{L} \sim \mathcal{N}(m_{\sigma_{L}}, sd_{\sigma_{L}}) \\
    \label{eq:prior_stress}
\end{equation}

\begin{equation}
    \mu_{\sigma_{L}} \sim \mathcal{N}(m = 50, sd = 5) \\
    \label{eq:prior_mean_stress}
\end{equation}

\begin{equation}
    \sigma_{\sigma_{L}} \sim LogNormal(m = 6, sd = 3) \\
    \label{eq:prior_SD_stress}
\end{equation}

\begin{equation}
    SCF, \sigma_{Y} \sim \Phi_{G} \Bigg( \begin{matrix} \Gamma(\alpha_{SCF}, \gamma_{SCF})\\ LogNormal(\mu_{\sigma_{Y}}, \sigma_{\sigma_{Y}}) \end{matrix}, \Sigma(SCF, \sigma_{Y}) \Bigg)
    \label{eq:prior_SCF_Strength}
\end{equation}

\begin{equation}
    \alpha_{SCF} \sim \mathcal{N} \bigg( m = 2, sd = \dfrac{1}{2} \bigg), \; \alpha_{SCF} \geq 0 \\
    \label{eq:prior_alpha_scf}
\end{equation}

\begin{equation}
    \gamma_{SCF} \sim \mathcal{N} \bigg( m = \dfrac{1}{2}, sd = \dfrac{1}{2} \bigg), \; \gamma_{SCF} \geq 0 \\
    \label{eq:prior_gamma_scf}
\end{equation}

\begin{equation}
    \mu_{\sigma_{Y}} \sim \mathcal{N}(m = 400, sd = 20) \\
    \label{eq:prior_mean_strength}
\end{equation}

\begin{equation}
    \sigma_{\sigma_Y} \sim LogNormal(m = 10, sd = 3) \\
    \label{eq:prior_SD_strength}
\end{equation}

\begin{equation}
    \Sigma(SCF, \sigma_{Y}) = \begin{bmatrix} Var[SCF] & \rho(SCF, \sigma_{Y}) \cdot \sqrt{Var[SCF]} \\ \rho(SCF, \sigma_{Y}) \cdot \sqrt{Var[SCF]} & \sigma_{\sigma_Y}^2
    \end{bmatrix}
    \label{eq:prior_SCF_strength_cov}
\end{equation}

\begin{equation}
    Var[SCF] = \alpha_{SCF} \times \gamma_{SCF}^2
    \label{eq:var_SCF}
\end{equation}

\begin{equation}
    \rho(SCF, \sigma_{Y}) = \dfrac{2}{3}
    \label{eq:corr_SCF_strength}
\end{equation}

An implementation of this decision analysis using the above models has been made available using the Julia programming language \cite{Bezanson2017} (with supporting optimisation libraries \cite{Salo2021, Lubin2022}). Samples from these priors have been obtained (using latin hypercube sampling \cite{Olsson2003}) for prior predictive checks, and solving the risk management decision problems \cite{DiFrancesco2023}. These samples are then used to estimate the expected value of combinations of material testing, SHM and inspection.

The utility function in this calculation is defined in Table \ref{tab:cost_fn}, providing the cost of implementing the various interventions is defined. Note that every combination of the risk mitigation actions in Table \ref{tab:bridge_actions} is considered in the calculation. A site visit cost (which is only incurred as part of a risk mitigation activity) is associated with each action. All costs in this example are normalised, so that the consequence of failure is set to $1.0$. Risk mitigation costs can therefore be considered as proportions of the failure cost.

\begin{table}[]
    \centering
    \begin{tabular}{|c|c|c|}
         \Xhline{2\arrayrulewidth}
        Intervention or outcome & Cost components & Total cost \\
         \Xhline{2\arrayrulewidth}
         repair &  \makecell{repair $(0.025)$ \\$+$ site visit $(0.01)$} & $0.035$ \\
         \hline
         replace &  \makecell{replace $(0.075)$ \\$+$ site visit $(0.01)$} & $0.085$ \\
         \hline
         reduce operation &  \makecell{reduce operation $(0.05)$ \\$+$ site visit $0.01$} & $0.05$ \\
         \hline
         repair and replace &  \makecell{repair $(0.025)$ \\$+$ replace $(0.075)$ \\$+$ site visit $(0.01)$} & $0.11$ \\
         \hline
         \makecell{repair, replace, \\and reduce operation} & \makecell{repair $(0.025)$ \\$+$ replace $(0.075)$ \\$+$ reduce operation $(0.05)$\\ $+$ site visit $(0.01)$} & $0.16$\\
         \hline
         failure & failure $(1.0)$ & $1.0$\\
         \hline
    \end{tabular}
    \caption{Evaluations of utility function for various combinations of interventions}
    \label{tab:cost_fn}
\end{table}

As an example, the expected utility associated with each risk mitigation option is evaluated using Equation \ref{eq:exp_cost}. This incorporates the cost of failure, $C_{fail}$ and of implementing any risk mitigation, $C_{rm}$. Note that maximising the expected utility (see Equations \ref{eq:prior_decision} and \ref{eq:prepost_decision}) is equivalent to minimising the expected cost. As shown in Table \ref{tab:prior_da}, the optimal action, $a^{*}$, conditional on the prior models of $\sigma_{Y}$, $SCF$, and $\sigma_{L}$ is to not invest in any risk mitigation measures in the upcoming window.

\begin{equation}
    E[u] = Pr(fail) \times - (C_{fail}) - C_{rm}
    \label{eq:exp_cost}
\end{equation}

\begin{table}[]
    \centering
    \begin{tabular}{|c|c|c|c|}
         \Xhline{2\arrayrulewidth}
        Intervention or outcome & $Pr(fail)$ & Risk mitigation cost(s) & Total cost \\
         \Xhline{2\arrayrulewidth}
         no action &  $0.0357$ & $0$ & \makecell{$a^{*}$ \\$0.0357$} \\
         \hline
         repair & $0.0101$ & $0.035$ & $0.0451$ \\
         \hline
         reduce operation & $0.0271$ & $0.050$ & $0.0771$ \\
         \hline
         replace & $0.000$ & $0.085$ & $0.085$ \\
         \hline
         \makecell{repair and \\reduce operation} & $0.0069$ & $0.085$ & $0.0919$ \\
         \hline
         repair and replace & $0.000$ & $0.110$ & $0.110$ \\
         \hline
         \makecell{repair, replace, \\and reduce operation}  & $0.000$ & $0.160$ & $0.160$ \\
         \hline
    \end{tabular}
    \caption{Evaluations of utility function for prior decision analysis (with no additional data collection)}
    \label{tab:prior_da}
\end{table}

The VoI analysis presented in Section \ref{sect:staff_bridge_results} compares this outcome to the results conditional on updated models of $\sigma_{Y}$, $SCF$, and $\sigma_{L}$, based on simulated (hypothesised) data from testing, inspection, and SHM.

\subsection{Results} \label{sect:staff_bridge_results}

\subsubsection{Perfect Information}

\begin{figure}
    \centering
    \includegraphics[width = \textwidth]{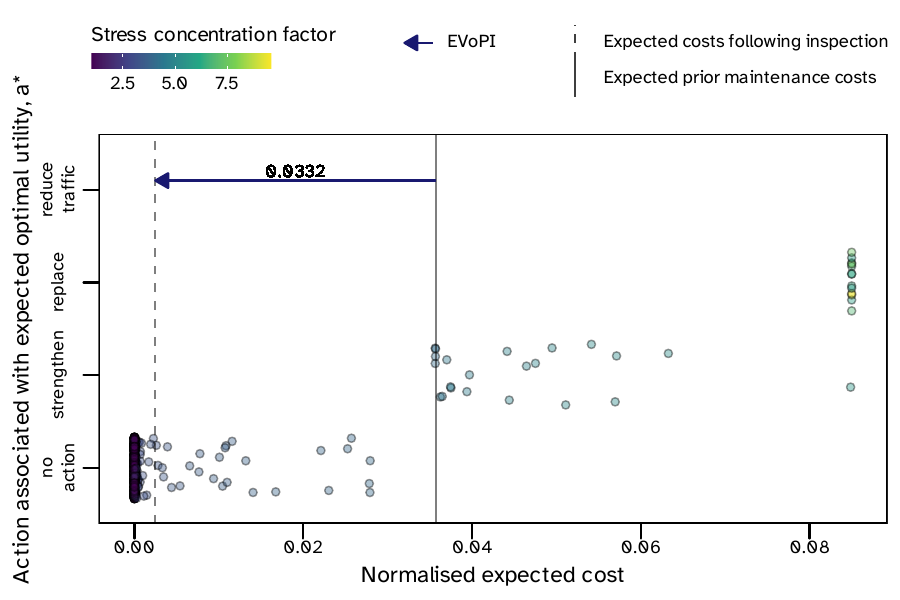}
    \caption{Expected value of perfect inspection data}
    \label{fig:VoPInsp}
\end{figure}

The expected value of perfect SHM data in the context of planning for an upcoming annual maintenance window is presented in Figure \ref{fig:VoPInsp}. This plot is the result of the decision analysis (expected optimal risk mitigation) associated with each sample from the prior model of stress. As the colorbar indicates, in instances where a relatively low stress concentration is inferred from the inspection data, the expected optimal solution is not to invest in maintenance activities. As the hypothesised misalignment measurement increases, so does the expected cost, initially because the associated higher stress concentration will expose the operator to more risk, but eventually then also because the expected optimal decision transitions to a strategy of investing in a strengthening repair to reduce the stress in the component. This continues until the strategy transitions again, this time to replace the component and remove any stress concentration effects due to poor installation. The mean value of the expected costs from all of these simulations is an estimate of the expected maintenance cost with perfect SHM data, and is indicated by the dashed vertical line. This is compared to the expected costs without this data, as indicated by the solid vertical line (results presented in Table \ref{tab:prior_da}). The arrow pointing to the difference between these lines is the expected value of the data.

\begin{figure}
    \centering
    \includegraphics[width = \textwidth]{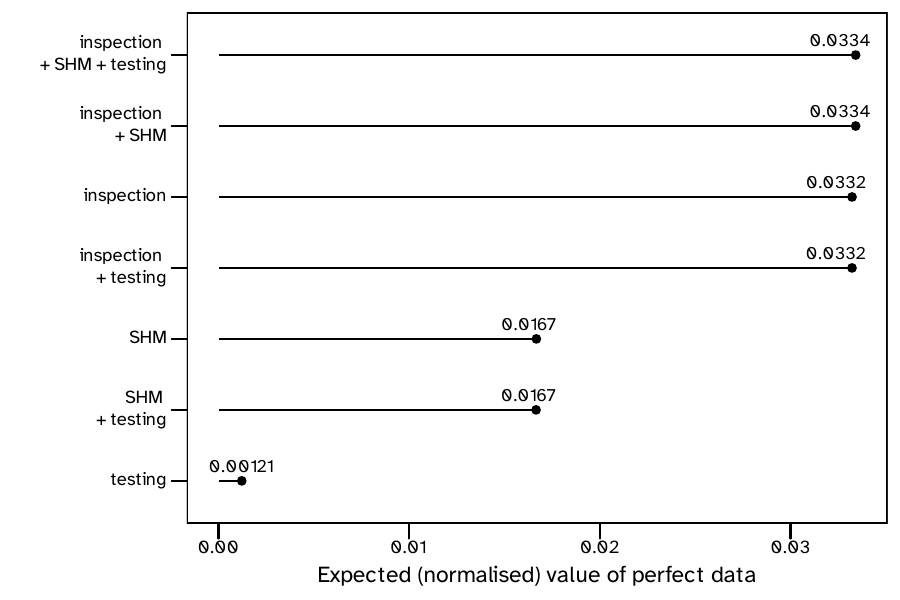}
    \caption{Expected value of combinations of structural integrity data}
    \label{fig:VoPI}
\end{figure}

This analysis can be repeated for inspection and testing, as well as combinations of these, i.e. assessing the influence of collecting multiple sources of data. These results are presented in Figure \ref{fig:VoPI}.

The key finding from this calculation is that these estimates do not simply sum linearly. For instance, the expected value of inspection, $VoPInsp$ was estimated to be $0.0332$ and the expected value of SHM, $VoPSHM$, was found to be $0.0167$. However, the expected value of collecting both types of data together was was estimated to be $0.0334$, which is less than $VoPInsp + VoPSHM$. Similarly, the expected value of testing alone was estimated to be $0.0012$, but it is not expected to contribute any further value when completed with either SHM or inspection (or both). This means that in instances where data will be collected from an inspection or SHM system, the maintenance strategy (and associated costs) is expected to be unchanged if material testing is also completed.

When a data collection is being evaluated the decision problem changes and the extent to which the new optimal action space now benefits from further reductions in uncertainty (from other types of data) may change.
These non-linearities are introduced by the utility function, for example due to the decision boundaries presented in Figure \ref{fig:VoPInsp}. This sometimes non-intuitive transformation onto a utility scale makes it difficult to generalise results, as the analysis is evidently dependent on the features of the specific decision problem. It does, however, introduce the benefit of producing interpretable, actionable results by providing an operator with a maximum (generally) monetary value that they should be willing to spend, for a specified type(s) of data. If a vendor quotes a higher price than the expected value of the data, then the risk optimal solution is simply to proceed without the data. What the results in Figure \ref{fig:VoPI} demonstrate is that when there is the opportunity to collect different types of data, these should be considered jointly in a VoI analysis, or a sub-optimal data collection plan may be identified.

\subsubsection{Imperfect Information}

\begin{figure}\centering
  \includegraphics{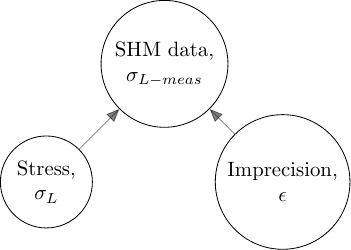}
  \caption{Simple DAG representation of measurement precision in SHM sensor data}
  \label{fig:SHM_prec}
\end{figure}

Quantifying the expected value of imperfect data increases the complexity of step 3b in Algorithm \ref{alg:VoI}. Here, a likelihood function is required to describe the imperfect features of the data, and the subsequent posterior distribution is then propagated through the decision analysis. For example, in the case of assessing the expected value of SHM, a measurement from the sensor no longer removes all uncertainty from the estimate of stress (at the time and location of measurement). Rather it is acknowledged that the the measurement may be an under-estimate, or over-estimate of the true stress. This imprecision can be easily incorporated into a Bayesian model, as shown in Figure \ref{fig:SHM_prec} and Equation \ref{eq:SHM_prec}. Note that as the precision with which the stress can be estimated from the SHM data increases, the expected value of this data asymptotically approaches the expected value of perfect SHM data, as calculated in Figure \ref{fig:VoPI}.

\begin{equation}
    \sigma_{L-meas} \sim N(\sigma_{L}, \epsilon)
    \label{eq:SHM_prec}
\end{equation}

\begin{figure}
    \centering
    \includegraphics[width = \textwidth]{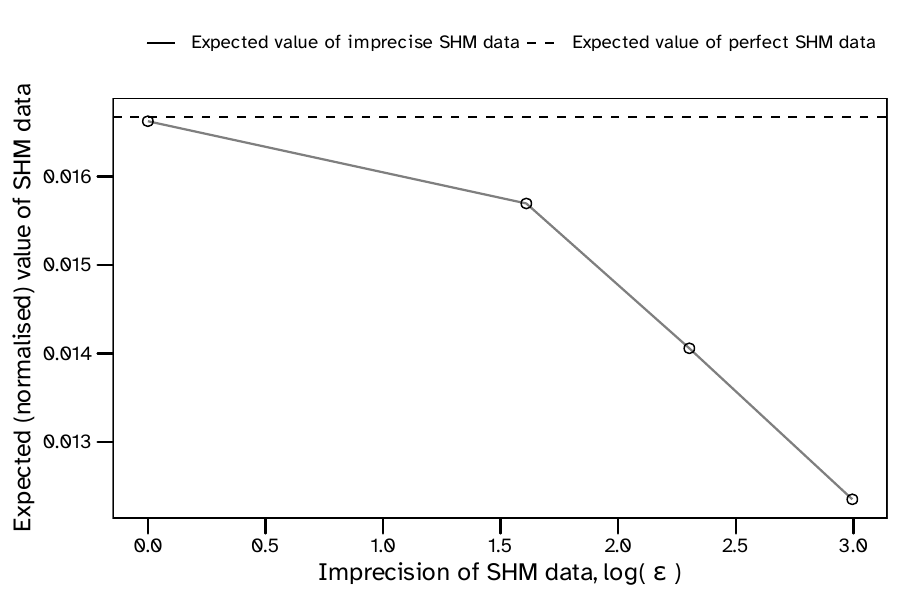}
    \caption{Sensitivity analysis of measurement uncertainty on expected value of SHM data}
    \label{fig:VoSHM}
\end{figure}

The expected value of perfect data provides a useful upper bound, and may be sufficient to support a decision (for instance if it is still below the cost that is being quoted to obtain the data, then it can be concluded that purchasing this data is not expected to be a risk optimal strategy). As data becomes increasingly imprecise, the extent to which it will support risk management decisions will decrease (or, at best, remain the same). This is demonstrated by the sensitivity analysis results in Figure \ref{fig:VoSHM}. Here, the value on the x-axis is $\epsilon$ in Equation \ref{eq:SHM_prec}. Note that this pattern is true for other features of imperfect information, such as bias, missingness, reliability, and risk of obtaining the data. Examples of how to incorporate these in a VoI analysis are provided in \cite{DiFrancesco2021}.

\subsubsection{Forecasting Multiple Maintenance Windows}

Another way to extend this problem is to quantify the expected value of data in the context of supporting multiple decision problems. Maintenance strategies are not static and depend on planned data collection, or risk mitigation interventions at other points in time. To address this, the influence diagram in Figure \ref{fig:rail_sim_id} can be repeated sequentially to represent the decisions that will need to be made in multiple future maintenance windows. This representation is generally referred to as a dynamic Bayesian network (or influence diagram) \cite{Morato2019}. This can be considered a system effect in the dependency between various elements in time, because they are jointly optimised. 

\begin{figure}
    \centering
    \includegraphics[width = \textwidth]{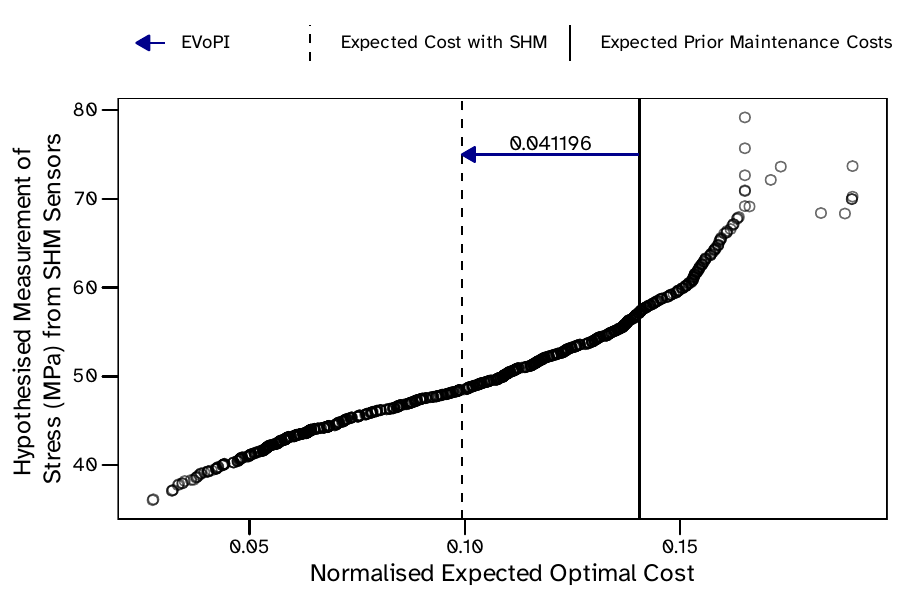}
    \caption{Prior samples of stress and expected optimal preposterior utilities for the repeated decision problem using three consecutive planning windows}
    \label{fig:my_utility}
\end{figure}

\begin{figure}
    \centering
    \includegraphics[width = \textwidth]{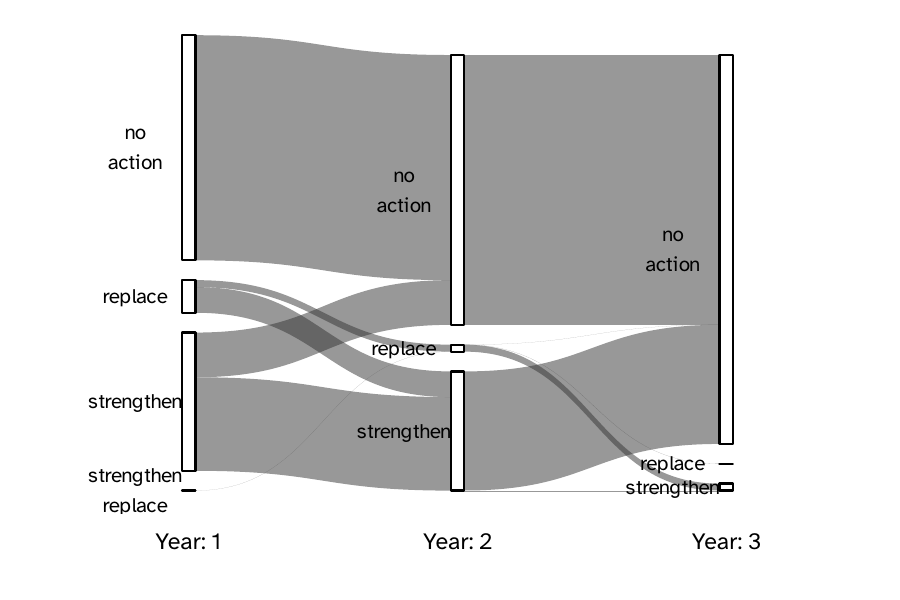}
    \caption{Sequences of optimal actions (w.r.t. expected utility) forecasting over a three-year period}
    \label{fig:my_joint}
\end{figure}

Solving the dynamic influence diagram provides the results in Table \ref{tab:prior_my_solutions}. However, when simulating (indirect) measurements of stress from an SHM sensor, the mean expected optimal cost again reduces and the expected VoI can be quantified as shown in Figure \ref{fig:my_utility}. Note that in each case, a specific combination of actions was identified, as was the case in the prior analysis in Table \ref{tab:prior_my_solutions}. The frequency of optimal action pathways is visualised in Figure \ref{fig:my_joint}. Here, the width of each path is proportional to the number of samples for which that action pathway minimises expected costs.

\begin{table}[]
    \centering
    \begin{tabular}{|c|c|c|c|}
         \Xhline{2\arrayrulewidth}
         $a^{*}$ Year 1 & $a^{*}$ Year 2 & $a^{*}$ Year 3 & $E[u(a^{*})]$ \\
         \Xhline{2\arrayrulewidth}
         strengthen & strengthen & no action & $0.155224$ \\
         \hline
    \end{tabular}
    \caption{Optimal actions (w.r.t. expected utility) forecast for three successive maintenance windows}
    \label{tab:prior_my_solutions}
\end{table}

\section{Conclusions} \label{sect:discussion}

The key conclusions and propositions from this work are outlined below:

\begin{itemize}
    \item Influence diagrams (or similar formalisation of structural integrity management decision problems) demonstrate the relationships/dependencies between different elements of a risk mitigation system. This includes showing which specific risks are mitigated by various interventions, and which quantities are measured by different data collection activities. Subject matter experts can agree this structure, which then lends itself to probabilistic modelling (uncertainty quantification) and optimisation (decision analysis) using modern software libraries. For instance, the calculation associated with the quantitative example in this paper is available in full in \cite{DiFrancesco2023}.

    \item Value of information analysis can be used to formalise the process of managing engineering data. This is contingent on a model, a mathematical description of the underlying decision problem, and how the (prospective) data is related to the quantities of interest. Ensuring (and transparently demonstrating) that sufficient data is obtained to effectively manage the risk of engineering systems aligns with the legal and ethical responsibilities of the profession. Training in methods of data-centric engineering, and improvements to industrial decision-support software are both expected to make VoI easier to engage with and unlock its benefits for professional engineers.

    \item The assessment procedure presented in this paper, namely
    \begin{enumerate}
    \item Describing the relationship between the quantities of interest with each other and prospective data collection \& risk mitigation actions.
    \item identifying expected optimal maintenance actions, and;
    \item quantifying the expected benefit of collecting additional data, in the context of further supporting this decision,
    \end{enumerate}
    is generic. If SHM technologies allow for other quantities to be measured with the same flexibility, then this approach will still be capable of finding the expected optimal combination of data to purchase.
    
    \item In this paper it has been demonstrated that when multiple measurement opportunities are present, failing to solve this problem jointly (considering inter-dependencies of collecting different combinations of various types of data) can lead to sub-optimal data collection plans, i.e. either gathering redundant information, or failing to collect information that becomes beneficial only in combination with other measurements.
    
\end{itemize}

\section{Acknowledgements}

Domenic Di Francesco is supported by the Ecosystem Leadership Award under the EPSRC Grant EP/X03870X/1, and The Alan Turing Institute, particularly the Turing Research Fellowship scheme under that grant.

Max Langtry is supported by the EPSRC, through the CDT in Future Infrastructure and Built Environment: Resilience in a Changing World Grant EP/S02302X/1.

Chris Dent was supported by the Isaac Newton Institute for Mathematical Sciences, in particular the Mathematical and Statistical Foundations of Data Driven Engineering programme when work on this paper was undertaken. This work was supported by the EPSRC Grant EP/R014604/1. He was also partially supported by a grant from the Simons Foundation.

\bibliography{refs}

\end{document}